\documentclass[aps,prb,twocolumn,showpacs,amsmath,amssymb,superscriptaddress]{revtex4}
\usepackage[english]{babel}
\usepackage{graphicx}
\usepackage{float}
\usepackage{amssymb,amsmath}
\usepackage{color}
\usepackage{textcomp}

\begin{document}

\title{Proton 
strings and rings 
in atypical nucleation 
of ferroelectricity in ice}

\author {J. Lasave} 
\affiliation{ Instituto de F{\'i}sica Rosario, CONICET and
	Universidad Nacional de Rosario, 27 de Febrero 210 Bis, 2000 Rosario,
	Argentina}
\affiliation{International Center for Theoretical Physics (ICTP), Strada
	Costiera 11, I-34151 Trieste, Italy}

\author{S. Koval}
\affiliation{ Instituto de F{\'i}sica Rosario, CONICET and
	Universidad Nacional de Rosario, 27 de Febrero 210 Bis, 2000 Rosario,
	Argentina}

\author{A. Laio}
\author{E. Tosatti}
\affiliation{International Center for Theoretical Physics (ICTP), Strada
Costiera 11, I-34151 Trieste, Italy}
\affiliation{International School for Advanced Studies (SISSA), and CNR-IOM
	Democritos, Via Bonomea 265, I-34136 Trieste, Italy}



\begin{abstract}

Ordinary ice has a proton-disordered phase which is kinetically metastable,
unable to reach spontaneously the ferroelectric (FE) ground state at low temperature
where a residual Pauling entropy persists. 
Upon light doping with KOH at low temperature the transition to FE ice takes place, but its microscopic mechanism still needs clarification. We introduce a lattice model based on dipolar interactions plus a competing, frustrating term that enforces the ice rule (IR). In the absence of IR-breaking defects, standard Monte Carlo (MC) simulation leaves this ice model stuck in a state of disordered proton ring configurations with the correct Pauling entropy. A replica-exchange accelerated MC sampling strategy succeeds, without open path moves, interfaces or off-lattice configurations, to equilibrate this defect-free ice, reaching its low-temperature FE order through a well defined first order phase transition. When proton vacancies mimicking the KOH impurities are planted into the IR-conserving lattice, they enable standard MC simulation to work, revealing the kinetics of evolution of ice from proton disorder to partial FE order below the transition temperature. Replacing ordinary nucleation, each impurity opens up a proton ring generating a linear string, an actual ferroelectric hydrogen-bond wire that expands with time. Reminiscent of those described for spin ice, these impurity-induced strings are proposed to exist in doped water ice too, where IRs are even stronger. The emerging mechanism yields a dependence of the long time FE order fraction upon dopant concentration, and upon quenching temperature, that compares favorably with that known in real life KOH doped ice.

\end{abstract}

\date{\today}
\maketitle

\section{Introduction}

Ice famously intrigues experimentalists and theoreticians alike. The
crystal structure of ordinary 
$I_h$ ice
consists of  hexagonal rings of
water molecules, each molecule tetrahedrally hydrogen-bonded to other four. Every
proton occupies one of two sites along the H-bonds between two oxygens, and
each oxygen satisfies the ice rule
(IR),
with two incoming and two ongoing
H-bonds {\cite{Ber33,Rah72}}. At low temperatures and under ordinary conditions, protons are unable to reach
equilibrium, and ice is
kinetically stuck in a glassy state characterized by the celebrated
Pauling entropy {\cite{Pet99,Pau35}} 
resulting from an
infinity of different, IR-conserving, defect free 
proton configurations. 
Still debated is the possibility to attain,
in pure ice at low $T$,
the low energy, zero-entropy ferroelectric (FE) phase, ice XI,   
endowed with 
a macroscopic polarization order parameter, the protons occupying bond sites in a unique order. 
An intimate understanding of the possible FE ordering 
mechanism
of pure ice is necessary in order to understand
why and how it is avoided or reached.  Of fundamental importance, that issue also bears 
potential implications in 
fields as disparate as
astrophysics {\cite{Bra99,Ume10}} and surface science {\cite{Su98,Sug16}}.

The well known extrinsic ingredient which experimentally permits realization of
ferroelectricity in bulk ice is KOH-doping, which 
allows ice to undergo, upon cooling
below $T_c$ $\approx$ 72K {\cite{Taj82}},  
the transition from proton disorder to proton-ordered 
ferroelectricity. This transition 
has several remarkable features. First, its
temperature is practically independent of the concentration and even of dopant
type, suggesting the FE phase and 
its onset
are in fact intrinsic equilibrium
features of ice, escaping realization merely due to kinetic reasons when doping is absent {\cite{Taj84}}.
Second, the ferroelectric
order achieved after long annealing times is partial and its fraction depends weakly
on the dopant concentration
in a wide range \cite{Fuk15}. Third, the transition kinetics upon quenching is significantly 
dependent on the quenching temperature $T_q$ {\cite{Fuk02,Fuk05}}. For example, at $T_q = 0.75 T_c$ 
no ordering is observed, but at \ $T_q = 0.89 T_c$ a fraction of the FE phase appears. 
As a simplifying note, we mention 
that nuclear quantum effects, generally relevant to hydrogen-bonded 
systems \cite{Kov02}, 
may be provisionally neglected
here, since deuteration affects only modestly the
transition temperature 
{\cite{Fuk05,Pam15}}.

The bulk of these observations is rationalized by 
the understanding
that an FE state of
low temperature hexagonal ice is thermodynamically favored, but its
realization is hindered by a kinetic slowdown, likely due to the IR. The slowdown is overcome only in special
conditions such as  doping, accompanied by 
quenching  
from a sufficiently high $T$ and 
subsequent annealing. 
Moreover, in many conditions the transition to the FE phase is incomplete, indicating that the slowdown mechanism can act also at the mesoscale.
To physically clarify this scenario, it is
desirable,
due to the complexity of the problem,
to resort to some simplified and yet microscopic 
model. The model should offer a 
comprehensive 
description of the non-ergodic proton disorder, of the 
ordered FE state, of their properties and ideal phase transition. It should also allow the introduction of dopants, with 
access to the  transformation kinetics to partial FE ordering which they permit. 

Microscopic model descriptions of ice are abundant, including off-lattice
descriptions, both force-field based, such as Ref. ~\cite{Don05} 
and ab initio. In particular, density functional theory (DFT) calculations also using graph invariants 
predict the  existence of an FE transition at $T_c$=98K in pure hexagonal ice {\cite{Sin05}}.
A better estimation of $T_c \approx$70-80K was obtained by recent DFT-based Monte
Carlo (MC) simulations using hybrid functionals {\cite{Sch14}}. An FE ordered phase
is also predicted in cubic ice by ab initio calculations, and is comparable in
energy to the corresponding FE phase of hexagonal ice {\cite{Raz11}}. A recent
study compares measured infrared spectra with theoretical results from
classical molecular dynamics and ab initio simulations suggesting evidence of
partial FE proton ordering in cubic ice {\cite{Gei14}}.
A possible FE transition at low $T$ in pure hexagonal ice was studied by lattice models using empirical
water potentials, yet with relatively inconclusive results owing to
strong dependence upon details of the potential models {\cite{Buc98,Ric05}}.
Alternatively, MC ice simulations using a point charge lattice model
led to a nonferroelectric state at low T {\cite{Bar93}} but it was argued that
the failure to find an FE state could be due to the simplifications of the
model. 
In addition, many specific observations have been made concerning the importance of rotational Bjerrum defects {\cite{Pet99}},
the role of Coulomb {\cite{Lek97, Lek98}}, of  multipolar interactions {\cite{Tri06}}, and other aspects {\cite{Kni06, Kni07}} 
including off-lattice hydroxide configurations {\cite{Cwi09}}. 

None of these 
realistic 
off-lattice concepts and considerations, however, seems  so far to lead to a well defined model description of the subtle kinetic phenomena
connected with establishing partial FE order, or the lack of it due to
persistent and glassy proton disorder, formidable problems that a simplified
lattice model is more likely be able to tackle.  That calls therefore for a fresh attempt.

We have developed a bare-bone lattice model of ice, and a MC technique which
allows simulating large samples at arbitrarily low temperatures (see {\it Calculation Details}).
The main
feature of this model is that while embodying "dipole-dipole" interactions,
its finite temperature ensemble contains only configurations
which satisfy exactly the IR. 
Even if a lower energy FE state is stabilized by dipolar interactions, 
standard MC 
fails to evolve configurations of this system even at finite temperature, 
causing it to retain the 
statistical distribution of proton rings and the
Pauling entropy 
down to
arbitrarily low temperatures, as   
in real neat ice. This ice-rule obeying disordered state provides an 
ideal framework where the effect on kinetics of an added idealized dopant can
be studied. 

The physics of our ice model, built on a diamond lattice, bears similarities to that of spin ice in pyrochlores, whose lattice
is dual to diamond, making the two isomorphic upon identification of the orientation of the spins with the location of hydrogen atoms on the bonds between oxygens \cite{Cas12}. 
Unlike spin ice models our model "water oxygens" possess interacting dipoles whereas the only interaction between hydrogens ("spins") come from the topological ice-rule constraints. These analogies and differences underpin those that will appear in the equilibrium phase diagram, reminiscent but not identical to the 3D Kasteleyn transition \cite{Kas63} of spin ice in a field \cite{Jau09,Fer16},  
as well as to the out-of-equilibrium, proton ordering behaviour. 
The kinetic process by which 
the dopant triggers proton ordering
is an avalanche of proton hoppings, which breaking up closed rings,  generate strings 
of collinear hydrogen bonds, all pointing in the same direction 
along a winding line,  thus upsetting
the ring landscape of the disordered phase.
The barrier characterizing this simple process, which is of the order of the dipolar interaction, is therefore
the rate limiting step for the formation of the string in ice. 

Before introducing the details of our work, it should be stressed that our model study 
omits,
by deliberate choice,
many details that are known to play a role in real ice. In spite of that, we shall nonetheless throughout the paper compare the model's
main results with known experimental facts. Points of agreement and disagreement between them will gauge the  model's ability to 
address mechanisms that underlie  some of the unexplained behaviors of ice ferroelectricity. 
In particular we will show that the string
formation qualitatively
reproduces
several known facts in real KOH doped ice, 
providing bare-bone mechanisms for 
the dependence of the ferroelectric fraction on the dopant molar
concentration, and on the quenching temperature.

\section{Model}

Our model system is a diamond lattice of $N$ "water molecules" where
each oxygen is connected tetrahedrally with four neighboring ones by H-bonds, as sketched
in Fig. \ref{Fig1}(a). This is the connectivity of cubic ice $I_c$. The cubic and  hexagonal ($I_h$) phases of ice differ 
by the stacking order of the
hexagonal bilayers that form the lattice, but their topologies 
are similar {\cite{Raz11}}. Since the model Hamiltonian, which is presented
below, depends on the connectivity of the lattice sites but not on the
distance between particles, the results obtained here should be valid
for both $I_c$ and $I_h$. 

Each oxygen at site $i$ in the diamond
lattice has four bonds to nearby sites labeled by $j$. The
variables of our model are the proton configurations on all bonds. We
represent them by a set of 4$N$ variables $\varphi_{ij}$, one for each {\it directed}
bond. 
We have $\varphi_{ij} = 1$ if in that bond there is a proton closer to oxygen $i$ and
$\varphi_{ij} = - 1$ if not. 
Note that in general $\varphi_{ji}$ is independent from  $\varphi_{ij}$.  In pure ice, where we
exclude Bjerrum defects \cite{Bje52}, 
all bonds 
possess one and only one proton,  and all oxygens two protons, $\varphi_{ij} = - \varphi_{ji}$, and the 
independent variable number shrinks to $2N$.
A bond Ising-type variable $\vec{\sigma}_i^j$ is defined 
as: $\vec{\sigma}_i^j = \varphi_{ij}  \vec{e}_{ij} $,
where $\vec{e}_{ij}$ 
is a vector pointing from oxygen site $i$ to site $j$.
Using these variables we define the dipole
associated with oxygen site $i$ as:
$ \vec{d}_i = \sum_j \varphi_{ij}  \vec{e}_{ij} = \sum_j \vec{\sigma}_i^j $. That definition corresponds to a dipole of modulus one in each site satisfying the IR
but is also valid for sites where it is not satisfied,  where the dipole moduli are now smaller than one. 

The
Hamiltonian of our model is
\begin{equation}
H = - J \sum_{(i, j)_{nn}} \vec{d}_i \cdot \vec{d}_j + k \sum_{i = 1}^N
(\sum_{j (i) = 1, \ldots 4} \varphi_{ij}
)^2 , \label{eq1}
\end{equation}
where the two control parameters, $J$ and $k$, are both positive. The first term
represents 
the  
nearest-neighbor (nn)
dipole-dipole 
ferroelectric interaction between oxygen tetrahedra. The second
term penalizes configurations that violate the IRs. Indeed, $\sum_j
\varphi_{ij} = 0$ only if two protons are close to site $i$ and two are far.

We will mostly describe the properties of this model, which to the best of our knowledge has not been studied for 
ice,  in the special
case $k \rightarrow \infty$, where violations of the IR are forbidden.
Yet, we will make use of finite $k$ in replica-exchange accelerated MC. The physical order parameter is the FE polarization, defined as

\begin{equation}
\vec{P} = \frac{1}{N} \sum_{i = 1}^N \vec{d}_i.
\end{equation}

The model can be mapped, translating from site to bond variables, to an Ising-type Hamiltonian  
\begin{equation}
H = - (J+k) \sum_{(il,jm)_{nn}} \vec{\sigma}_i^l \cdot \vec{\sigma}_j^m - 
J  \sum_{(il,jm)_{nnn}} \vec{\sigma}_i^l \cdot \vec{\sigma}_j^m. 
\label{eq2}
\end{equation}

\noindent The first term is a large (practically infinite) ferroelectric coupling between nearest neighbor bonds, promoting 
frustration and disorder through its strong  topological IR constraints.
That effect is mitigated by the second term, also ferroelectric,
between second neighbor bonds, contributing instead to stabilize a possible FE ordered state at low T.

It should be noted that off-lattice configurations \cite{Cwi09} 
as well as multipolar terms and 
long-range interactions 
\cite{Lek97, Lek98, Tri06, Kni06, Kni07}, are omitted. 
Testing the effects of removing these drastic approximations is beyond the scopes of this first study. Mainly justified by simplicity, 
the short-range interaction assumption is at least encouraged by screening of long-range electrostatic tails, which is induced by polarization. It may also be noted that,
unlike first neighbor interactions, always ferroelectric, the sign of long-range interactions is not uniform, but 
rather oscillates between ferro and antiferro depending on direction, suggesting a certain level of cancellation. Indeed, 
MC studies of the dipolar spin ice model 
actually showed 
that medium to long range 
interactions are screened  
out, suggesting that short range physics should remain qualitatively valid \cite{Mel04,Her00,Bra01,Cas12}.

The KOH impurities, which play a fundamental role in determining the kinetics, are introduced in our model as follows. In ice the $K^+$ impurity replaces a  proton in one bond.  
This  turns the proton-deficient molecule into a cation-hydroxide pair $K^+(OH)^-$. 
That is simulated in our model by a single, fixed proton vacancy in a  bond $ij$ (represented by setting $\varphi_{ij} =\varphi_{ji} = -1$), an action which  simultaneously deprives oxygen site $i$ of an outgoing proton, (this is the Bjerrum defect mimicking $K^+$,
which we keep fixed),  and deprives oxygen site $j$ by one incoming proton -- this is a mobile IR breaking defect, mimicking ($OH)^-$.   We have no charges in our model, but for the sake of illustration, we will call these two defects  $K^+$ and $OH^-$.
The presence of charge dopants in real doped ice induces lattice distortion, which could lower the relaxation barrier of the local structure, speeding the interconversion from paraelectric (PE) to FE. 
However, the formulation of our model does not allow assessing how important this particular effect is.

\section{Calculation details}

We carried out MC simulations on a diamond lattice with 12x12x12
cubic cells containing 
$N=$ 13824 sites,
representative of static oxygen sites in cubic ice. 
A z-directed electric field
is coupled to the polarization for breaking the symmetry of the isotropic Hamiltonian. The field is
removed after equilibration and is small enough
($|\vec{E}| \approx J/10$)
in order not to modify the transition temperature.
First of all we confirmed that 
the IR term causes  
frustration that prevents our defect free ice 
model from reaching thermodynamic equilibrium 
within standard 
MC sampling, where 
proton
variables $\varphi_{ij}$
change one at a time
while  chosen randomly through the lattice.
$2N$ proton move attempts performed sequentially represent our MC step or pass.
This well known problem was addressed long ago by Rahman and Stillinger {\cite{Rah72}} who dealt with the IR by performing random walks on the lattice and noticing that paths involving crossing of the periodic boundaries bring in a change in total dipole moment. Time-honored as it is, that method involves the necessity of very large simulation sizes, which we prefer to avoid. 
We thermalize the system by a Hamiltonian Replica Exchange Method (HREM) {\cite{Bun00}}. We simulated $m = 1, 2, ...M$ 
replicas in parallel, at the same temperature and same $J$ but with different values of the ice-rule penalty parameter $k$.  
The original replica $m = 1$ has a  
prohibitively 
large value of $k=30J$, 
practically infinite.
As $m$ increases the parameter $k$ is reduced successively until the last replica 
$m = M = 40$, which 
corresponds to $k = 0$.
The IR-violating defects (excess or lack of protons attached to an oxygen site) will
therefore occur with increasing probability in replicas with decreasing $k$.
After a prescribed number of MC steps the instantaneous configurations for
adjacent replica are allowed to swap with a probability dictated by the
standard Metropolis exchange criterion {\cite{Aff06,Huk96}}.

As a direct extension of the intrinsic, defect free model, whose equilibrium properties 
will be shown to agree well with those of pure ice, we 
subsequently 
introduced defects, and studied 
the kinetics which they generate.
This is done in a lattice of
9x9x9
cubic cells containing 5832 sites.
To represent the effect of "KOH type" 
impurities,
a small number of
$L \ll N$ fixed
proton vacancies were introduced,
randomly distributed in lattice bonds
$ij$, 
by setting
$\varphi_{ij} = \varphi_{ji} = - 1$, i.e. no proton either near oxygen $i$ or near oxygen  $j$,
as if it had been moved to the nearby interstice in order to mimic the role of K$^+$ \cite{Pet99}.
In Fig. \ref{Fig3} the
proton vacancy in a given bond is depicted by an interstitial K$^+$ ion
schematically replacing the proton H$^+$.
In real ice, the KOH impurity produces two mobile defects:
an ionic OH$^{-}$ defect and a Bjerrum $L-$defect (proton vacancy) \cite{Pet99}.
In our model, for the sake of simplicity, the Bjerrum $L-$defect
is fixed and only the hydroxide defect is able to move.
Thus, the introduced impurity generates a traveling hydroxide which can trigger transitions involving the
nearby protons, transitions otherwise impossible
(see Fig. \ref{Fig3}(b)). 
It's worth noting here that we expect similar
kinetic effects from a mobile Bjerrum $L-$defect as those observed with a traveling hydroxide.

To address the kinetics of the doped ice model, we performed non-equilibrium simulations by 
a quenching-annealing (QA) simulation protocol, carried out with standard MC moves -- which 
mimicks to some extent the real time evolution of experiments -- and by comparison also 
with the HREM protocol, which artificially speeds up evolution towards equilibrium. 
Each calculation performed was an average of 50
runs with different random-number generator seeds.
Starting with the doped 
system initially thermalized 
with HREM
at very high T $\sim 3$ T$_c$, we quench it down to a temperature 
T$_q$ below T$_c$ (quenching) and then let it thermalize at the quenching temperature 
T$_q$ till the system comes as close as possible to equilibrium (the annealing process). That was
done for a range of  
T$_q$
and of doping concentrations, so as to address the known
experimental dependence of ice ferroelectricity upon these parameters.

\section{Results}

\subsection{Equilibrium phase diagram and proton rings}

\begin{figure*}
	\includegraphics[scale=0.22]{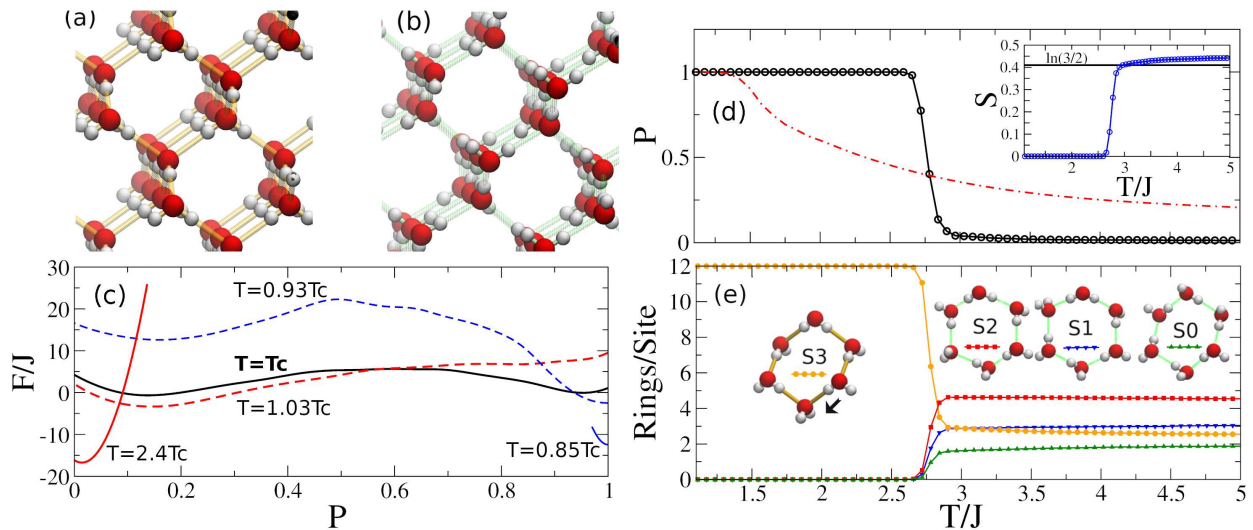} 
	\caption{(a) Simulation snapshot showing the FE order in ice at $T < T_c$.
		(b) Simulation snapshot showing a characteristic PE configuration in ice at
		$T > T_c$.
		(c) Free energy $F$ in units of the coupling constant $J$ as a function of polarization $P$ for different temperatures. 
		The FE ($P \sim 1$) and PE  ($P \sim 0$) minima at $T_c$
		are separated by a barrier, consistent with a first order transition.
		(d) Equilibrium averaged polarization $P$ vs temperature  
		in units of $J$ obtained by HREM
		for the model of Eq. \ref{eq1} (black
		solid line and empty circles). 
		Red dotted-dashed line:  
		Kasteleyn-like polarization as a function of 
		$T/|\vec{E}|$ for  
		a spin-ice model ~\cite{Goh19} with the same IRs as 
		in Eq. \ref{eq1}, but $J=0$ and with $z$-oriented electric-field  $\vec{E}$
		coupled to dipoles ($-\vec{d}.\vec{E}$).
		Inset: entropy in units of the Boltzmann constant vs $T/J$ for our ice model (blue line and empty 
		circles). The horizontal solid line 
		indicates the value of the  
		Pauling entropy, 
		which actually persists down to low $T$ in ordinary MC simulations.
		(e) Hexagonal ring population vs
		temperature in units of $J$. The insets show schematically the proton arrangements for the
		different type of rings: S$_0$, S$_1$, S$_2$ and
		S$_3$. We also show the arrow 
		associated to a given configuration of an O-H-O bond in the ring S$_3$, which is defined for computing the ring-order parameter $s$ (see {\it Equilibrium Phase Diagram and Proton Rings}).
	}
	\label{Fig1}
\end{figure*}

We first discuss the equilibrium phase diagram for defect-free bulk ice model, where no violations to the IR are allowed. In order to thermalize this model we use a 
Hamiltonian Replica Exchange Method (HREM)
(see {\it Calculation Details}), in which a set of replicas differing only by the IR-controlling parameter $k$, are simulated in parallel. In the 
first replica $k$ is extremely large, as appropriate to the IR conserving model we want to address, in the other replicas $k$ is succesively smaller and smaller.
Configurations of different replicas are exchanged according to a replica exchange protocol, which  allows the simultaneous thermalization of all the replicas. 
Fig. \ref{Fig1}(d), shows the average value of the polarization $P$ as a function
of temperature. There is a transition between an ordered FE state ($P \sim 1$) and a disordered 
PE state ($P \sim 0$) at an equilibrium transition temperature $T_c \approx 3 J$. 
As expected, $T_c$ is proportional to the strength
of the oxygen dipolar-interaction parameter $J$ (see Eq. 
\ref{eq1}).
The transition appears to be strongly first order. 
Even without the  IR constraint ($k$=0), the symmetry-dictated universality class of this transition 
would differ from straight Ising.   
Indeed, the Hamiltonian (Eq. \ref{eq1})
possesses six equivalent FE ground states (polarization along x, -x; y, -y; z, -z), making it closer (yet not identical) 
to a Potts model,  a family many members of which support  first order phase transitions  in high dimensions (see e.g. Ref.~\cite{Wu82}). 
Conversely, the fully IR conserving Hamiltonian would, once the dipole-dipole interaction was removed ($J$=0)
and the protons were coupled to an electric field,
display a Kasteleyn-type transition \cite{Kas63} as  
in spin ices.
With nonzero $J$ and large $k$, our model is richer,  even if retaining some qualitative similarities to Potts and Kasteleyn transitions. 
A note of caution here is that while the experimental 
FE transition 
of real ice is, as in this model, first order {\cite{Pet99, Bra99}}, there are 
in ice secondary order parameters, such as strain coupling, that are absent in the model but that would play a role making the transition first order.

Next, the equilibrium entropy evolution with temperature is a crucial information. We obtain it at each temperature 
as the integral of the specific heat at constant volume over $T$.
Strictly speaking, this procedure is correct only if no first-order phase transitions are encountered along the path. However, in finite size systems like those analyzed in this work, first order transitions are avoided, the thermodynamic potentials vary continuously, and the procedure is therefore justified.
The inset
of Fig. \ref{Fig1}(d)
shows how at $T_c$ the 
entropy of the
defect free IR conserving model  
correctly rises from 
essentially
zero (the model has no acoustical modes) to the Pauling value 
$S \sim $  ln(3/2) 
across the transition.
The free energy $F$ as a function of the polarization can be estimated, at a given temperature, from the histogram of the polarization  
observed in the first replica.
$F$ is shown in Fig. \ref{Fig1}(c). At $T_c$, $F$ shows two
minima with same
free energy separated by a barrier, which confirms the 
first-order character of the transition.
At $T=0.93T_c$ the free energy retains a secondary minimum at $P \sim 0$, signaling a metastable (equilibrated) PE state which, however, no longer exists at $T=0.85T_c$. An analogous metastable FE state must also exist above $T_c$, but is already lost at $T \sim 1.03T_c$. Thus, free-energy barriers vanish
shortly below and just above $T_c$.
Associated with the transition there is a change in the proton configuration inside the 12 hexagonal rings 
which thread each lattice site. The role of rings and directed H-bonds 
is widely discussed in ice and water \cite{Don05,Has13}.   
Here, we must
in addition
distinguish 
different ring types
according to their polarization.  
For that,
we associate an arrow to each H-bond, pointing from the oxygen 
possessing a  
close-by proton  
to the other oxygen
in that bond (see Fig. \ref{Fig1}(e)). We then count, for each ring, the number of arrows  pointing in a specific 
clockwise direction,
and define from that a directed order parameter $s= ( 6 - |\sum_{i=1}^6 \varphi_{il}| ) / 2  $, where $i$ runs through 
the six-site ring 
clockwise, $l=i+1$, and $\varphi_{il}$ is the proton variable of the H-bond $il$.  
The four different 
kinds of proton rings are schematically depicted in Fig. \ref{Fig1}(e)
labeled as $S_{s}$, therefore $S_3$, $S_2$, $S_1$, and $S_0$, corresponding to  $s=3, 2, 1$ and $0$, respectively. This ring
classification differs from a previous one (see Supp. Inf. of Ref.~ \cite{Has13}), except for the case of the ring $S_0$.

A schematic representation of a typical microscopic configuration in the
PE phase of ice is depicted in Fig. \ref{Fig1}(b). 
Inspecting all hexagonal rings in the equilibrium state of the defect-free ice model, we extract the average population $<n_{s}>$ of rings
of each $s =0,..3$. Fig. \ref{Fig1}(e) shows the results obtained as a function of temperature.  
For each site
$\sum_{s}^3 <n_{s}> = 12$ in pure ice at all temperatures.
Well below $T_c$, all rings have $s=3$, thus $n_3 = 12$, accompanied by a net local polarization along $z$. 
In this FE state, schematically displayed in Fig. \ref{Fig1}(a), where only z-polarized $S_3$ rings are present, the symmetry 
between the 
six possible polarizations of the system (along the x, y or z axis, and
corresponding negative directions)
is spontaneously broken by long-range
order.
At $T_c$, $<n_3>$ has a sharp drop  
which accompanies 
the collapse of the order parameter $P$,
while all other rings  
concurrently 
surge and
proliferate as shown in Fig. \ref{Fig1}(e).
Finally, in the PE
phase above $T_c$, all ring populations acquire steady values, almost constant 
with  further temperature increase. The $S_3$ rings do not disappear, but we find them equally polarized in all directions in accordance with the vanishing order parameter. In the following, $S_2$, $S_1$, and $S_0$ are called "disordered" rings because they only appear in the disordered phase.
The slight residual temperature dependence can be attributed to finite size in our simulations. 

Entropy reveals another effect of small size.  Our calculated entropy
at $T  >  T_c$  is $\approx$ $10\%$  higher than the Pauling entropy as shown in
the inset of 
Fig. \ref{Fig1}(d).
Pauling's  entropy is known to be only a lower bound \cite{Fer16}. Our HREM  calculations capture
the additional
proton correlations along the closed rings, which cause entropy to rise higher for smaller sizes \cite{Her14}. 

The population distribution of rings in the disordered phase is  also similar to that found in a
recent ab initio molecular dynamics study of hexagonal ice \cite{Has13}.
For instance, the relative abundance of $S_0$ rings is about 15.8\% in our calculation which
is nearly equal to the corresponding averaged-value obtained in 
Ref.~\cite{Has13} for hexagonal ice, $\approx$ 16.5\%.

\subsection{Non-equilibrium kinetics 
and FE polarization 
in doped ice}

Thus far we described the static 
properties, both equilibrium and 
metastable,  of the ice model. 
It is now possible to  address the nonequilibrium MC kinetics of transformation between 
PE and FE states.
The equilibrium transition being first order, 
the transformation
will occur by nucleation. Yet,  this process is very strongly influenced by IR constraints,
which render ordinary
homogeneous nucleation impossible, at least for $k$ large enough.  In that limit, only inhomogeneous nucleation is possible.  
We therefore study 
the transformation from the metastable and proton disordered state, 
into an ordered or partly ordered FE state, 
taking place
once  
model impurities, meant to  play a similar role to 
KOH, are introduced.   
To that end, we 
first equilibrate 
with HREM
the IR conserving state at high
$T \sim 3T_c$, 
and then quench it
down to some $T_q$ below $T_c$ where we let it thermalize 
with standard MC moves.
We call this procedure a quenching-annealing (QA) simulation
protocol (see {\it Calculation Details}).

\begin{figure*}
	\includegraphics[scale=0.72]{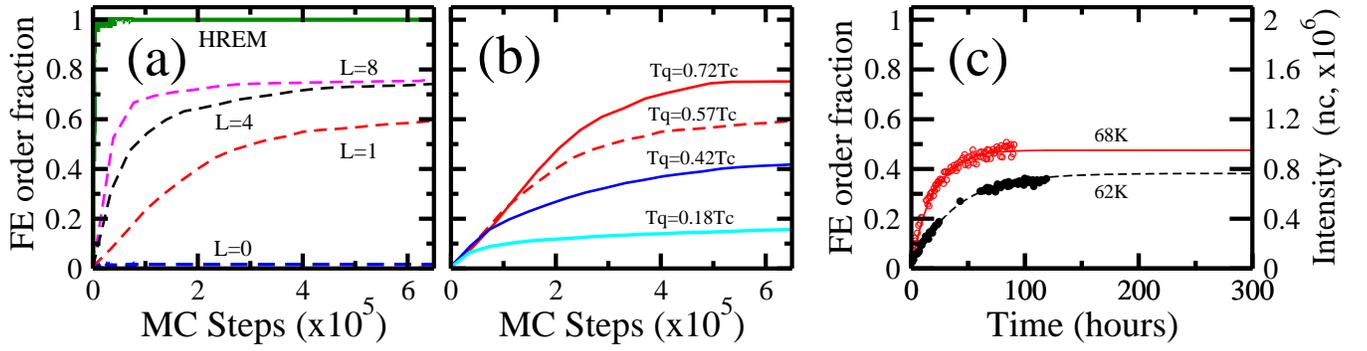}
	\caption{(a), (b) FE order fraction (or equivalently, instantaneous polarization) 
		attained after quenching vs 
		the number of MC steps.   
		(a) FE fraction 
		for a quenching temperature $T_q = 0.57 T_c$
		and different number $L$ of impurities.  For $L =$ 0, we show with blue
		dashed line the regular MC results and with green solid line the HREM
		results. Regular MC results for $L =$ 1, 4, and 8 (molar fractions 1/5832, 4/5832 and 8/5832, respectively) are shown with red dashed
		line, black dashed line and magenta dashed line, respectively. (b) Regular MC
		results for $L = 1$ (molar fraction 1/5832) and different quenching temperatures. Results for $T_q =
		0.18 T_c$, $0.42 T_c$, $0.57 T_c$, and $0.72 T_c$ are shown with turquoise
		solid line, blue solid line, red dashed line, and red solid line,
		respectively. (c) Evolution with time of the
		131-Bragg peak  
		neutron count (nc, right ordinate) obtained in  
		diffraction experiments of KOD-doped deuterated ice {\cite{Fuk02}}. Solid black (open red) circles are from the same
		sample once
		annealed at $T=$ 62 (68) K. 
		Left ordinate:  
		corresponding ice-XI mass  fraction (a measure of FE order, proportional to Bragg intensity) 
		for the sample annealed at $T=$ 62 K (black dashed line) and 68 K (red solid line), after Ref.{\cite{Fuk02}}.
	}
	\label{Fig2}
\end{figure*}

\begin{figure*}
	{\resizebox{7.0 in}{!}{\includegraphics{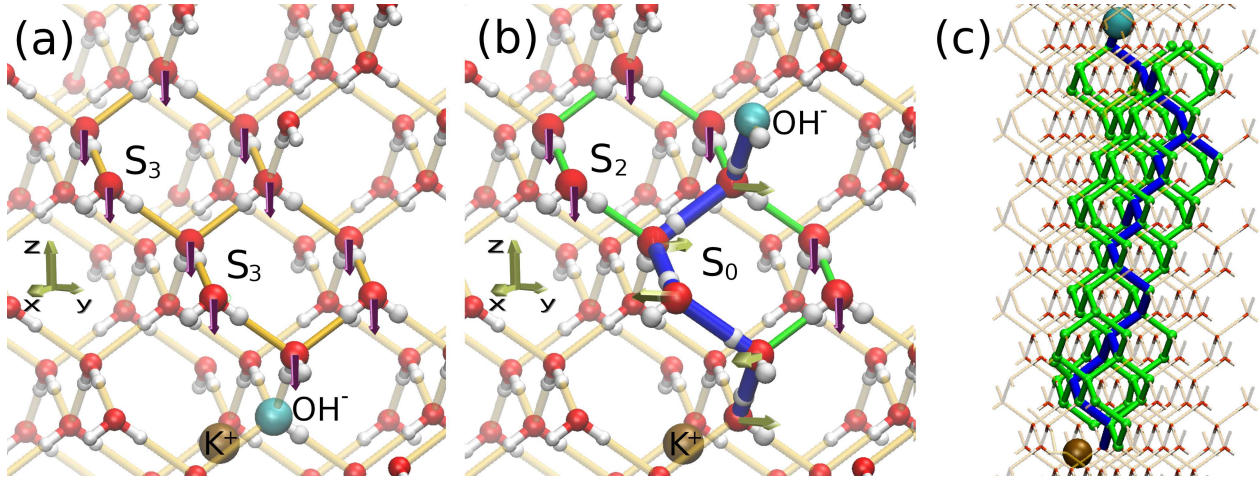}}}
	\caption{Nucleation mechanism of disorder in the ordered FE phase of ice
		depicted with snapshots of the simulation at different stages of the process
		(see explanations in {\it Microscopic Mechanism of String Nucleation}): (a) in the doped FE phase at $T = T_c / 2$ with
		the characteristic S$_3$ rings colored with mustard. The planted
		proton vacancy representing the doping by a KOH impurity produces a
		hydroxide defect (colored with turquoise) which remains in its site at this
		simulation temperature. Notice that we added a fixed K$^+$ atom
		colored with brown next to the proton vacancy for the sake of clarity in the
		picture as explained in {\it Calculation Details} (see also {\it Microscopic Mechanism of String Nucleation}); (b) immediately after the suddenly raise of T above
		$T_c$ showing the hydroxide displacement through single proton
		jumps following the blue path and transforming two z-polarized mustard rings
		S$_3$ into two ``disordered'' green rings S$_0$ and
		S$_2$; (c) after a longer simulation time  
		above $T_c$
		showing the creation of a PE cluster (green rings) along the blue path of
		the hydroxide. Violet (green) arrows at different oxygens represent
		z-polarized (xy-polarized) dipoles $ \vec{d}_i$.  }
	\label{Fig3}
\end{figure*}

Initially, after a certain number of 
thermalization time steps 
at high $T$, the system reaches
a state where all the ionic
defects 
(hydronium - hydroxide pairs and even molecular
states with zero or four protons) 
introduced 
by the random initial configurations
managed to recombine and disappear. We checked that after thermalization
at high $T$ in the KOH-doped system with $L$ extrinsic impurities, we have
precisely $L$ 
mobile hydroxide defects
in the system because all intrinsic ionic
defects permitted by finite $k$ have recombined
(see Fig. \ref{Fig3}).
After equilibration at high T with HREM, all the simulations continue with a QA protocol using standard MC (unless otherwise stated) on the replica with the largest $k$ value.
Fig. \ref{Fig2}(a) shows the non-equilibrium evolution of the instantaneous polarization in 
a QA simulation  
after quenching at $T_q = 0.57T_c$, in
a range of different conditions. 
As a first check,  
in pure ice ($L = 0$) the system remains stuck in a non-equilibrium glassy state with $P \ll 1$  as shown by the blue dashed curve
in Fig. \ref{Fig2}(a). 
If HREM is instead kept active 
throughout, then
the low-temperature thermodynamic equilibrium with 
$P = 1$ (FE order) is quickly recovered
after quenching,
as expected and as shown by the green curve
in Fig. \ref{Fig2}(a).  

The next step is the
simulation of
doped ice 
with $L$ impurities representing KOH  
impurities
(see \textit{Model}
and \textit{Calculation Details}).
Unlike the undoped case ($L = 0$), results for $L =$ 1, 4, and 8, show a kinetic
evolution with 
frank
onset of
the FE order parameter (see Fig. \ref{Fig2}(a)). At 
large MC step number (conventionally representing long evolution times), 
the polarization 
$P$ reaches  
$\approx$ 60 $-$ 75\%,
almost independent of the impurity number $L$. This is in qualitative agreement with neutron
diffraction measurements of doped deuterated ice, where a volume abundance of $\approx$ 48\% of ice XI is observed in the bulk at
$T=0.89T^{exp}_c$ {\cite{Fuk02}}. The critical temperature and the ice XI  fraction locally formed are practically
independent of the impurity concentration {\cite{Taj82,Taj84,Tya02,Fuk05,Fuk15}},
a nontrivial outcome which 
is reproduced by
our model.
The $L = 1$ case corresponds to a molar fraction of 1/5832, similar to that
of the doped-ice samples used in Ref.\cite{Taj84}, 1/5540, and leads to extensive FE ordering in the
model that is qualitatively similar to experiment {\cite{Taj82,Taj84,Fuk02}.
	Moreover, neutron diffraction of annealed 
	KOD-doped deuterated ice
	after low $T$
	quenching showed a sustained intensity growth of the characteristic 131-Bragg
	peak of the FE phase XI. Its intensity,  
	proportional to the volume fraction of 
	ferroelectric ice-XI,  
	tends to a definite limit at long annealing times {\cite{Fuk02}}, 
	also decreasing when the quench temperature was lowered, as shown in
	Fig. \ref{Fig2}(c). 
	We conducted 
	additional extensive
	weak-doping simulations, with $L = 1$,  exploring
	how a change of $T_q$ affects the kinetics of FE onset. As Fig. \ref{Fig2}(b) shows, the
	calculated long-time FE fraction diminishes as the quenching temperature is lowered, in qualitative agreement 
	with the neutron diffraction data of Fig. \ref{Fig2}(c). 
	There are therefore good hopes that our model could shed light on the underlying reasons.

\subsection{Microscopic mechanism of 
string 
nucleation} 

We now analyze the microscopic mechanism of
IR defect-induced disorder nucleation in the FE phase and conversely, 
the impurity-induced nucleation and growth of  
FE clusters in the PE phase. First, we address the 
impurity-triggered nucleation of disorder by regular MC simulations of an ordered FE crystal at $T = T_c / 2$. 
At this low temperature, 
as shown in Fig. \ref{Fig3}(a),
the hydroxide is unable to propagate freely through the crystal.
However, when 
temperature rises above T$_c$ ($\approx 2 T_c$)}, 
the hydroxide  
departs from the 
impurity site, and 
travels through the lattice 
(see Fig. \ref{Fig3}(b)). In its journey,
it flips onto the xy plane
the dipoles
from their original FE z-polarization. That generates a chain, or string, of xy-dipoles, 
shown by the blue line in Fig. \ref{Fig3}(b), with
origin in 
the fixed initial impurity site and end at the moving hydroxide. 
This chain bears a resemblance to the so-called "Dirac" string associated with a
magnetic monopole 
of model 
spin ice systems {\cite{Cas08,Gin09, Mor09}}.
While the IR is of course the topological constraint that water ice and spin ice have in common
which gives rise to strings in both cases, the two model systems are far from identical, as we will underline later.

As shown in Figs. \ref{Fig3}(a) and \ref{Fig3}(b), in the early stages of string formation the
hydroxide can only progress upwards (see the blue path) along $z$ 
and against the total polarization. 
As the ice
rules are satisfied everywhere and the system is in the ordered phase, it can
only receive one of the two protons from the top neighbouring water molecules. 
The preference will be to receive the
one that creates locally a basal dipole aligned to 
that of
the previous step in the hydroxide 
path, thus creating a chain of dipoles
aligned in the x or y directions,
no longer along z. This has an energy cost of $\Delta E = 2 J$
per step.
Otherwise, the  
resulting
basal dipole 
would be
perpendicular to that of the previous step  
with a higher cost $\Delta E = 3 J$. 
Yet, since MC moves are randomly generated and accepted according 
to Boltzmann weights, the chain-end hydroxide   
progresses, owing to finite temperature, 
not only in the z direction but
also 
in the x or y directions, as in Figs. \ref{Fig3}(b) and
\ref{Fig3}(c).
As noted, the disordering mechanism produced 
onto the initially perfect FE system
by the hydroxide  string 
with a probability
that bifurcates
at each step along the path 
resembles 
that of
the three-dimensional Kasteleyn-like transition of spin ice in external field {\cite{Jau08}}. 
However, in our ice model the transition is not induced
globally by an external field, but by the 
$J$-induced 
local field created by the growing seed.
Unlike Kasteleyn's strict case of spin ice models, where an infinite number of configurations are degenerate and excitations have a large gap \cite{Cas12},  
here nucleation brings the system closer to a lower 
free energy state due to the local dipole-dipole interactions. 
A second difference is that the local dipole field makes the 
probability to move a proton in the two possible branches 
uneven, affecting qualitatively the nucleation dynamics.  

\begin{figure*}
	{\resizebox{7.0
			in}{!}{\includegraphics{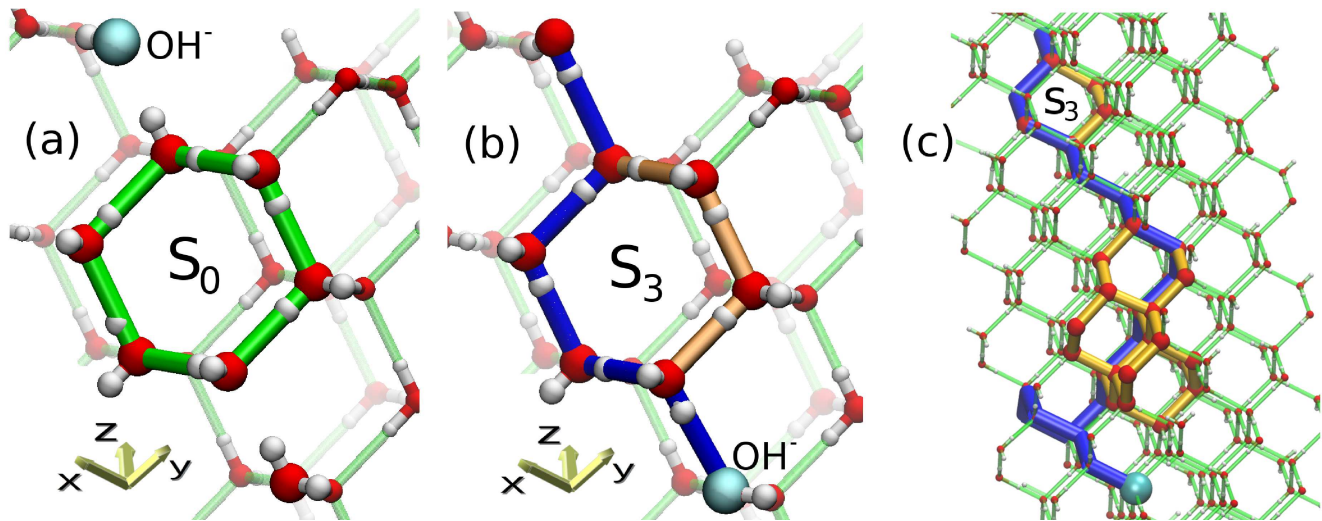}}}
	\caption{Nucleation mechanism of FE clusters inside the PE phase of ice
		depicted with snapshots of the simulation at different stages of the process
		(see explanations in {\it Microscopic Mechanism of String Nucleation}): (a) After thermalization at $T = 3 T_c$ showing
		the hydroxide defect (colored with turquoise) and different ``disordered''
		green rings; (b) immediately after quenching  
		below $T_c$
		showing the blue path of the hydroxide and the consequent conversion of a
		(green) S$_0$ ring into a z-polarized S$_3$ (mustard)  
		ring;  (c) after a longer simulation time at 
		$T < T_c$ showing the blue
		path of the hydroxide. In its travel, the hydroxide nucleates a new mustard
		S$_3$ cluster (shown at the top of the image) and enables to
		enlarge an already existing FE mustard cluster (shown at the bottom of the
		image).}
	\label{Fig4}
\end{figure*}

The disorder produced by the hydroxide migration may also be characterized 
by the transformation of z-polarized S$_3$ 
(mustard) rings into S$_2$, S$_1$ or S$_0$ (green)
``disordered'' rings with a certain degree of xy-polarization (see also Fig.
\ref{Fig1}). For instance, Fig. \ref{Fig3}(b) shows the formation of an
S$_0$ ring with a three-step hydroxide jump, and 
that of 
an S$_2$ ring
with a single-step jump. Thus, the traveling hydroxide nucleates in its path a
disorder "contagion" cloud,
formed by green rings,
elongated in the z direction
and zigzagging in the x and y directions, as shown in Fig. \ref{Fig3}(c).
The green cluster shown in this figure has a substantially
smaller polarization than the FE bulk, and  
can thus be considered a seed of the PE phase
inside the FE bulk. 
It is worth noting here that after the hydroxide has passed,
the green PE cluster cannot further spread expanding
its frontier perpendicularly
to the blue line because 
the strong IR constraints frustrate
any proton move attemp across the cluster boundary during the standard MC simulation,
as they presumably would in real time evolution.
In other words, the PE cluster can only progress
as an elongated string through the traveling-hydroxide tip.

In reverse, and crucially,
we finally address the 
nucleation
mechanism of FE  
order
inside the disordered phase. Thermalizing the system with a 
single impurity ($L=1$) at a high $T = 3 T_c$, with an initial HREM MC simulation 
lasting 20000 steps, 
we choose
the replica with the largest $k$ value, therefore with well-respected IRs. With that, a regular 
(Metropolis) MC 
simulation is continued for the same number of steps.
In this thermalization, the hydroxide migrates
following
a completely random path
and
losing track of the 
initial impurity site.
Fig. \ref{Fig4}(a) 
shows a typical 
configuration
formed after thermalization. Completely disordered, it displays all
types of S$_{\beta}$ rings as described earlier. 
This disordered configuration is then suddenly quenched  
to a low temperature below
$T_c$ (see Figs. \ref{Fig4}(b) and \ref{Fig4}(c)). The reduced 
mobility 
of the hydroxide 
and the decrease of entropic contributions in
favor of enthalpic ones 
reflects in the tendency of the hydroxide
to migrate preferentially in 
one direction, that will in fact define the
incipient polarization direction which we denote as $z$, 
as in Figs. \ref{Fig4}(b) and \ref{Fig4}(c). In its way along the new path, the
hydroxide transforms disordered green rings into ordered z-polarized mustard
ones. For instance, Figs. \ref{Fig4}(a) and
\ref{Fig4}(b) show the transformation of a S$_0$  
into a
z-polarized S$_3$ ring as the hydroxide progresses along the blue path.
This is precisely the reverse process to that displayed in Figs.
\ref{Fig3}(a) and \ref{Fig3}(b). In its journey, the hydroxide may also take
some steps 
that do not transform
disordered rings into  S$_3$ ordered ones, changing 
disordered rings into other disordered ones. 
Alternatively, this quenched  
evolution
may also enlarge an existing ordered cluster by expanding its frontier, where again disordered rings
turn into S$_3$ ones.
All three possibilities were observed and appear in the snapshot taken from the
simulation of Fig. \ref{Fig4}(c). The net total result is the nucleation
of ordered mustard rings along the blue path of the hydroxide.

\section{Discussion}

We have described how FE and PE states transform into one another in a bare bone lattice model of ice,
where only IRs and near-neighbor dipolar interactions are retained. 

This model, it should be clear, has no ambition of describing real water ice in all chemical details, a field in itself whose literature is immense.
The model however, is amenable to solution by simulation; and that makes it, as is often the case,
quite instructive. 

First we find, by means of an adequate MC 
protocol, that there is an equilibrium first order phase transition between the two states, with the correct Pauling entropy jump and
an instructive proton ring distribution in the FE and PE states. The equilibrium transformation between the two does not take place 
by regular nucleation as in normal first order transitions because, as  is known for a very long time, IRs make ordinary 
nucleation \cite{Kas00}  ineffective:  leaving pure, defect free bulk ice in a metastable PE state endowed by Pauling's entropy 
and a very characteristic proton ring distribution down to the lowest temperatures.  

By introducing  impurities, mimicking dopants such as KOH known experimentally to nucleate 
the transition, we examine the very special FE-PE and PE-FE heterogeneous nucleation mechanism in an IR-obeying system .  
The dopant generates an itinerant 
hydroxide-induced
defect whose string-like evolution inside the bulk effectively punctures, 
as it were, the otherwise infinite  barrier between the two states, ending the kinetic invulnerability of the metastable PE state at low temperatures.
Starting with the PE state, the growth of the hydroxide string provides a quasi one-dimensional heterogeneous nucleation
mechanism, with a propagating winding cloud of FE rings inside the initially proton-disordered bulk. 
This is in turn reflected by the increase of the FE order parameter  as  time (in our case MC time)  evolves after quenching, as 
simulations show (Fig. \ref{Fig2}(a)). Snapshots in Fig.
\ref{Fig4} (and Fig. \ref{Fig3})
moreover show a predicted  
FE (PE) nucleation
landscape  proceeding along the
string of flipped protons which acts as the backbone.
These strings and in fact the qualitative nature of the nucleation process 
are reminiscent 
of Kasteleyn-like transitions
in spin ice models-- not surprisingly, because the ordered (disordered) phase onset is 
again IR-dominated \cite{Cas12}.
Nonetheless, the differences are important. Already at equilibrium, Fig. \ref{Fig1}(d) compares the 
temperature dependence  of the order parameter of the ice model with that of a spin ice model 
(see Gohlke {\it et al.}~\cite{Goh19}) with the same IRs 
but with an external field instead of our local dipole-dipole interaction (i.e., with $J=0$ in our language). In spin ice there is a 3D Kasteleyn transition in the low-field regime with a 
characteristic second-order
behavior at $T > T_c$,  in contrast to the first-order behavior of our phase transition.
Beyond that, the evolution kinetics of strings in our ice model is controlled by $J$, again an element absent in spin ice models.

A number of results suggested by the present lattice ice model encouragingly resemble those known  either experimentally or in more elaborate off-lattice models of 
water ice. 

The static structure and ring correlations and the correct Pauling entropy of clean bulk ice appear to describe well the disordered PE state, as summarized by Fig. \ref{Fig1}.
The  capability of metal hydroxide dopants to give rise to growing FE strings inside the PE state and viceversa -- thus functioning as unconventional inhomogeneous 
nucleation agents -- is demonstrated, as in 
Figs. \ref{Fig3} and \ref{Fig4}.  
The long-time FE polarization fraction grows as the quenching temperature increases approaching $T_c$ (Fig. \ref{Fig2}(b)), in nontrivial agreement with neutron diffraction experiments and contrary to usual ferrodistortive structural transitions where clusters with reversed order parameters below $T_c$ lead to a decrease of the average order parameter as $T$ increases approaching the transition \cite{Sch78,Sch76,Yuk91}.
Again similar to  real ice, the dependence of FE polarization upon the dopant concentration is minimal. In the model, inhomogeneous nucleation occurs with any 
number of extrinsic centers, and  the residual increasing effectiveness appears simply to reflect a speed-up kinetics once the system is close to the transition point.
Finally, the slowing down in the growth rate of FE polarization order parameter with MC time also resembles that observed in water ice in real time-- see Fig. \ref{Fig2}.
In that slowing down 
however a   
multiplicity of elements can be simultaneously at work.  The  string nuclei cannot, owing to their nanoscale transverse size, 
convert, in a system with strict IRs, a PE state to complete  ferroelectricity. Probably even more important in practice, string nuclei might suffer a decrease of their growth rate when their tips hit other existing FE clusters, or, in real ice, grain boundaries and other lattice defects. In our simulations, that kind of effect is involuntarily introduced by finite size.
Even ignoring these important realistic aspects, one could note that the partially polarized system free energy is progressively closer to the FE equilibium state than the disordered starting point,  yielding a  decreasing thermodynamic force felt by the growing string tip ends.

Beyond purely on-lattice models like ours, the moving hydroxide can, besides moving on in-lattice configurations {\cite{ Kni06, Kni07}}, also visit (and be arrested by) off-lattice interstitial configurations {\cite{Cwi09}}. That event, not described in our model, will slow down the hydroxide mobility and also introduce kinks with possible bifurcations in the  
strings evolution.
An evolution which nevertheless our model depicts in its most elementary form.  

In conclusion, we have presented a soluble lattice model depicting the onset of ice ferroelectricity as a first order phase transition,  and demonstrating how nucleation and growth mechanisms,  otherwise universal in the kinetics of first order phase transitions, are profoundly changed by topological ice-rule constraints that control proton ordering. Results sheds light on, and support further understanding of, the onset and demise of ferroelectricity in ice.

\begin{acknowledgments}

E.T. thanks R. Car and S. Singer for very helpful discussions and inputs. J.L. and S.K. acknowledge fruitful discussions with S. Scandolo. J.L. also thanks A. Hassanali for helpful suggestions.
J.L. and S.K. acknowledge support from Consejo Nacional de Investigaciones Cient\'ificas y T\'ecnicas (CONICET), Argentina. E.T. is supported by ERC Advanced Grant N. 8344023 ULTRADISS, and in part by the Italian Ministry of University and Research through PRIN UTFROM N. 20178PZCB5.

\end{acknowledgments}

\bibliography{ice}

\end{document}